\documentclass[twocolumn]{aastex63}

\usepackage{amsmath}
\usepackage{graphicx}
\usepackage{wrapfig}
\usepackage{csvsimple}

\usepackage{appendix}
\usepackage{comment}
\usepackage{cprotect}
\usepackage{array, ltablex, multirow}
\usepackage{listings}
\usepackage{enumitem}
\usepackage{varwidth}
\usepackage{tasks}

\newcommand{\isofont}[1]{{\mathrm{#1}}}
\newcommand{\isomass}[1]{{\ensuremath{\isofont{^{#1}}}}}
\newcommand{\isocharge}[1]{{\ensuremath{\isofont{_{#1}}}}}
\newcommand{\isotope}[3]{{\ensuremath{\isocharge{#1}\isomass{#2}\isofont{#3}}}}

\newcommand{\I}[2]{{\isotope{}{#1}{#2}}}

\newcommand{\Ep}[1]{{\ensuremath{10^{#1}}}}
\newcommand{\E}[1]{{\ensuremath{\powersep\Ep{#1}}}}

\newcommand{\powersep}{\times}

\newcommand{\unitstyle}[1]{\ensuremath{\mathrm{#1}}}

\newcommand{\Kelvin}{\unitstyle{K}}
\newcommand{\K}{\Kelvin}  

\newcommand{\Msun}{\ensuremath{\unitstyle{M}_\odot}}
\newcommand{\Lsun}{\ensuremath{\unitstyle{L}_{\odot}}}

\newcommand{\code}[1]{\texttt{#1}}
\newcommand{\mesa}{\code{MESA}}
\newcommand{\MESA}{\mesa}

\newcommand{\GYRE}{\code{GYRE}}

\input{vectors.tex}

\newcommand{\D}{{\mathrm d}}

\newcommand{\nuclei}[2]{\ensuremath{\mathrm{^{#1}#2}}}

\newcommand{\COrate}{\ensuremath{\mathrm{^{12}C(\alpha, \gamma)^{16}O}}}

\newcommand{\hydrogen}[1][1]{\nuclei{#1}{H}}
\newcommand{\helium}[1][4]{\nuclei{#1}{He}}

\newcommand{\carbon}[1][12]{\nuclei{#1}{C}}
\newcommand{\nitrogen}[1][14]{\nuclei{#1}{N}}
\newcommand{\oxygen}[1][16]{\nuclei{#1}{O}}

\newcommand{\neon}[1][20]{\nuclei{#1}{Ne}}

\newcommand{\iron}[1][56]{\nuclei{#1}{Fe}}

\newcommand{\grada}{\ensuremath{\nabla_{{\rm ad}}}}

\newcommand{\gradT}{\ensuremath{\nabla_T}}
\newcommand{\Teff}{\ensuremath{T_{\rm eff}}}	

\newcommand{\BV}{Brunt-V\"{a}is\"{a}l\"{a}}

\newcommand{\chiT}{\chi_{T}}
\newcommand{\chir}{\chi_{\rho}}

\makeatletter
\newcommand*{\rom}[1]{\expandafter\@slowromancap\romannumeral #1@}
\makeatother

\newlength{\apjcolwidth}
\newlength{\figwidth}
\newlength{\doublewide}
\setlist[itemize]{noitemsep, topsep=0pt, leftmargin=*}

\begin{document}

\title{On Trapped Modes In Variable White Dwarfs As Probes Of The $^{12}$C($\alpha, \gamma$)$^{16}$O Reaction Rate}

\shorttitle{Trapped Modes In Variable WD as probes of $^{12}$C($\alpha, \gamma$)$^{16}$O}
\shortauthors{Chidester, Farag, \& Timmes} 

\author[0000-0002-5107-8639]{Morgan T. Chidester}
\affiliation{School of Earth and Space Exploration, Arizona State University, Tempe, AZ 85287, USA}
\affiliation{Joint Institute for Nuclear Astrophysics - Center for the Evolution of the Elements, USA}

\author[0000-0002-5794-4286]{Ebraheem Farag}
\affiliation{School of Earth and Space Exploration, Arizona State University, Tempe, AZ 85287, USA}
\affiliation{Joint Institute for Nuclear Astrophysics - Center for the Evolution of the Elements, USA}

\author[0000-0002-0474-159X]{F.X.~Timmes}
\affiliation{School of Earth and Space Exploration, Arizona State University, Tempe, AZ 85287, USA}
\affiliation{Joint Institute for Nuclear Astrophysics - Center for the Evolution of the Elements, USA}

\correspondingauthor{Morgan T. Chidester}
\email{taylormorgan32@gmail.com}


\begin{abstract}
We seek signatures of the current experimental $^{12}$C$(\alpha,\gamma)^{16}$O reaction rate probability distribution function
in the pulsation periods of carbon-oxygen white dwarf models. We find that adiabatic g-modes 
trapped by the interior carbon-rich layer offer potentially useful signatures 
of this reaction rate probability distribution function. 
Probing the carbon-rich region is relevant because it forms 
during the evolution of low-mass stars under radiative helium burning conditions,
mitigating the impact of convective mixing processes. 
We make direct quantitative connections between the pulsation periods of the identified trapped g-modes
in variable WD models 
and the current experimental $^{12}$C$(\alpha,\gamma)^{16}$O  reaction rate probability distribution function. 
We find an average spread in relative period shifts of $\Delta P/P \simeq \pm$\,2\%  for the identified trapped g-modes 
over the $\pm$\,3$\sigma$ uncertainty in the 
$^{12}$C$(\alpha,\gamma)^{16}$O reaction rate probability distribution function ---
across the effective temperature range of observed DAV and DBV white dwarfs
and for different white dwarf masses, helium shell masses, and hydrogen shell masses. 
The g-mode pulsation periods of observed white dwarfs are typically given to 6-7 significant figures of precision.
This suggests that an astrophysical constraint on the $^{12}$C$(\alpha,\gamma)^{16}$O reaction rate could,
in principle, be extractable from the period spectrum of observed variable white dwarfs.
\end{abstract}


\section{Introduction}\label{sec:intro}

Tens of thousands of nuclear reactions can participate during the evolution
of a star, depending on the environmental conditions.
Only a few of these reactions have a strong impact on the overall
chemical evolution of the elements, with significant consequences for
the chemistry and the subsequent molecular evolution of baryonic matter. 
In particular, the helium burning $^{12}$C($\alpha$,$\gamma$)$^{16}$O reaction
plays a major role in the energy production and nucleosynthesis of stars 
\citep[e.g.,][]{iben_1967_aa, fowler_1984_ab,arnett_1996_aa,iliadis_2015_ab,deboer_2017_aa}
and thus influences the $^{12}$C/$^{16}$O ratio in the universe.

The difficulty in measuring the $^{12}$C($\alpha$,$\gamma$)$^{16}$O
rate in nuclear experiments is due to the small cross section of the
$^{12}$C($\alpha$,$\gamma$)$^{16}$O reaction at temperatures relevant
for helium burning in stars.  Nuclear experiments in terrestrial
laboratories provide data for energies as low as about 2 MeV, with
extrapolation to stellar conditions at $kT$ $\simeq$ 20 keV.  At
stellar conditions, two partial waves contribute, denoted by their
spectroscopic E1 and E2 amplitudes in reaction and scattering theory
\citep{fowler_1984_aa}.  The challenge is measurement of the low energy
angular distributions of the $^{12}$C($\alpha$,$\gamma$)$^{16}$O
reaction, from which the E1 and E2 cross sections are extracted.

Decreasing the uncertainty in the $^{12}$C($\alpha$,$\gamma$)$^{16}$O
reaction rate from low-energy nuclear experiments has markedly
improved in recent years.  For example, to obtain a comprehensive
evaluation \citet{deboer_2017_aa} considered the entirety of existing
experimental data, aggregating about 60 years of experimental data
consisting of more than 50 independent experimental studies. More than
10,000 data points were then incorporated into a complete
multi-channel phenomenological R-matrix analysis.  A main result was
the characterization of the uncertainty in the reaction rate, which
was accomplished through a Monte-Carlo uncertainty analysis of the
data and the extrapolation to low-energy using the R-matrix model.
After finding an approximately Gaussian underlying probability
distribution for the rate, there was statistical significance with the
1$\sigma$ uncertainty of the reaction rate. 
A goal of forthcoming experiments is
to further reduce the uncertainty in the 
$^{12}$C($\alpha$,$\gamma$)$^{16}$O reaction rate \citep[e.g., ][]{smith_2021_aa}.

Partnering with this laboratory astrophysics quest are astrophysical
constraints on the $^{12}$C($\alpha$,$\gamma$)$^{16}$O reaction
rate. For example, \citet{weaver_1993_aa} find a permissible range of
the reaction rate by requiring the integrated nucleosynthesis yields,
from a set of massive star explosion models over plausible initial mass functions,
agree with the observed solar abundances for the
intermediate mass isotopes.  They found this range was insensitive to
the assumed slope of the initial mass function within observational
limits, and relatively insensitive to some details of convective boundary mixing.

As another example of an astrophysical constraint, models for the
evolution of single stars predicts the existence of a gap in the black
hole mass distribution for high mass stars 
due to the high temperature effects of electron-positron
pair production \citep{heger_2002_aa}.  
The location of the black hole mass gap is generally robust with respect to model
uncertainties \citep{takahashi_2018_aa, farmer_2019_aa, marchant_2020_aa}, 
but depends sensitively on the uncertain $^{12}$C($\alpha$,$\gamma$)$^{16}$O 
reaction rate \citep{farmer_2020_aa}.  
The location of the black hole mass gap, probed though LIGO/Virgo/Karga
\citep{ligo-scientific-collaboration_2015_aa, acernese_2015_aa, akutsu_2021_aa} 
gravitational wave determinations of the masses and spins of merging binary black holes, 
thus allows a constraint on the $^{12}$C($\alpha$,$\gamma$)$^{16}$O reaction rate 
\citep{farmer_2020_aa, woosley_2021_aa, mehta_2022_aa}.

Seismology of hydrogen dominated atmosphere (DA class) and helium dominated atmosphere (DB class) 
carbon-oxygen white dwarfs (CO WDs) has also been used to place constraints on the 
$^{12}$C($\alpha$,$\gamma$)$^{16}$O reaction rate
\citep{metcalfe_2001_aa,metcalfe_2002_aa}.
In this approach the central \oxygen\ abundance and the location 
of the oxygen-to-carbon transition within the CO core are used as fitting parameters when 
minimizing the difference between the observed pulsations periods of specific WDs
and the pulsation periods of WD models.  
The derived parameters can then imply a constraint on the
$^{12}$C($\alpha$,$\gamma$)$^{16}$O reaction rate, but are 
sensitive to model choices \citep{metcalfe_2003_aa},
diffusion \citep{fontaine_2002_aa}, and convective boundary mixing \citep{straniero_2003_aa}.

\citet{de-geronimo_2017_aa} analyzed two CO WD models with masses 0.548 and 0.837 \Msun\ derived 
from evolutionary calculations from the zero-age main sequence (ZAMS). They considered models 
that varied the number of thermal pulses, the amount of overshooting, and the
$^{12}$C($\alpha$,$\gamma$)$^{16}$O reaction rate within the uncertainties known at the time.
They found that independent variations of these quantities produced significant changes 
in the resulting DA WD chemical profiles and the pulsation period spectrum.

\citet{pepper_2022_aa} calculated evolutionary models with initial masses in the range of
0.90$\le$\,$M/\Msun$\,$\le$3.05. 
They considered different $^{12}$C($\alpha$,$\gamma$)$^{16}$O reaction
rates within the uncertainties known at the time.  As expected, they
found no changes in the evolution prior to the core He-burning stage.
However, they found that the subsequent stages of evolution produced
differences in the convective He core mass, the number of thermal
pulses during the asymptotic giant branch phase of evolution, and
broad trends in the chemical profiles.

The main novelty of this article is a new search for potential signatures of the
current experimental $^{12}$C$(\alpha,\gamma)^{16}$O reaction rate probability distribution function 
in the pulsation periods of CO WD models.
Section \ref{sec:models} describes our models, 
Section \ref{sec:results} describes our results, 
Section \ref{sec:sensitivities} discusses sensitivities, and 
Section \ref{sec:summary} summarizes the results of our new search.

\section{Models}\label{sec:models}

\subsection{Updated $^{12}$C($\alpha,\gamma$)$^{16}$O reaction rates}\label{sec:rates}

\citet{mehta_2022_aa} expanded the \citet{deboer_2017_aa} 
tabulated reaction rate for $^{12}$C($\alpha,\gamma$)$^{16}$O  
to a finer temperature grid, from 52 to 2015 temperature points, 
to ensure that no temperature step results in variations in the rate of more 
than a factor of two. The recalculations also provided the
formal $\pm$\,3\,$\sigma$ uncertainties on the experimental reaction rate 
probability distribution function in steps of 0.5$\sigma$.
$\sigma_0$ is the median rate consistent with an astrophysical S-factor of S(300 keV)\,=\,140\,keV\,b
with a $\pm$1\,$\sigma$\,=\,21\,keV\,b uncertainty. By exploring $\pm$3\,$\sigma$\, we effectively 
explore the range S(300 keV) = (77,203)\,keV\,b, where positive and negative $\sigma$ indicate a stronger
and weaker rate than the median value, respectively.

\begin{figure*}[!ht]
    \centering
    \includegraphics[trim = {0.0cm 0.0cm 0.0cm 0.0cm },clip,width=\doublewide]{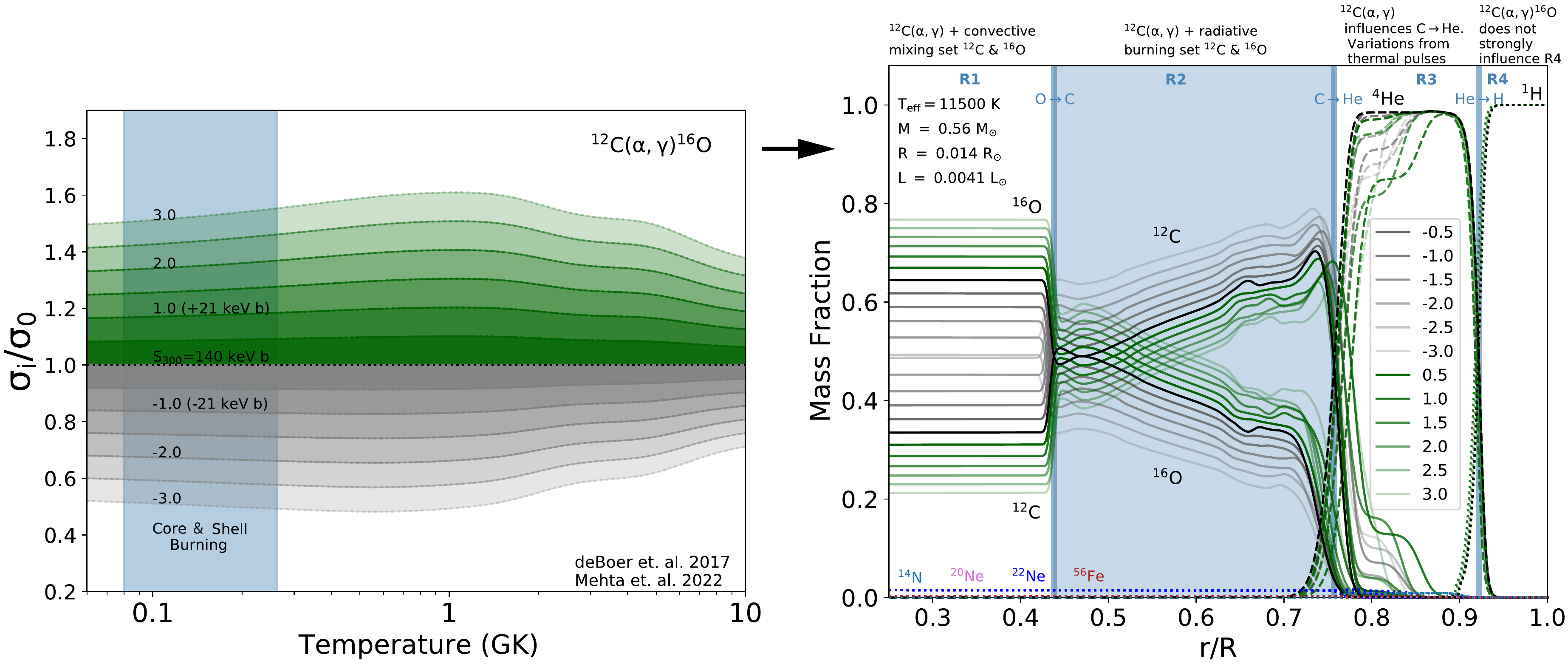}
    \caption{{\it Left:} $^{12}$C($\alpha,\gamma$) reaction rate ratios, $\sigma_i/\sigma_0$, as a function of temperature.
For our models, $\sigma_i$ spans -3.0 to 3.0 in 0.5 step increments, with $\sigma_0$ being the current nominal rate.  Negative $\sigma_i$ are gray curves and positive $\sigma_i$ are green curves. The $\pm$\,1,\,2,\,3 $\sigma_i$ curves are labeled.  The blue band show the range of temperatures encountered during core and shell He burning.
{\it Right:} Mass fraction profiles of the evolutionary DAV models resulting from the $^{12}$C($\alpha,\gamma$) reaction rate uncertainties $\sigma_i$ after each model has cooled to \Teff=11,500\,K. The nominal $\sigma$\,=\,0 reaction rate is the black curve, 
negative $\sigma_i$ are gray curves and positive $\sigma_i$ are green curves. Solid curves are for \carbon\ and \oxygen, dashed curves are for \hydrogen\ and \helium. The trace isotopes \nitrogen, \neon, \neon[22], and \iron[56] are labeled.  Key regions and transitions are also labeled (see text).}
    \label{fig:rate_ratios}
\end{figure*}

This probability distribution function is shown 
in Figure~\ref{fig:rate_ratios} over the $\pm$\,3\,$\sigma$ region.
Green bands represent positive
$\sigma_i$ while the gray bands represent negative $\sigma_i$.  
The 13 individual $\sigma$ curves are depicted by respective dotted lines,
with the $\pm{1,2,3} \sigma$ lines labeled.  The blue region shows the
temperature range spanned by core and shell He burning 
in our evolutionary 2.1\,M$_\odot$ ZAMS mass models.  
These reaction rate files are available at 
doi:\dataset[10.5281/zenodo.6779983]{https://doi.org/10.5281/zenodo.6779983}.

\subsection{MESA, wd\_builder, and GYRE}\label{sec:tools}

We use \MESA\ version r15140
\citep{paxton_2011_aa,paxton_2013_aa,paxton_2015_aa,paxton_2018_aa,paxton_2019_aa}
to evolve $2.1\,\Msun$, $Z$\,=\,0.0151 metallicity, non-rotating models
from the ZAMS to the top of the WD cooling track in the
Hertzsprung-Russell diagram.  This results in 0.56\,\Msun\ DA WDs
with $\simeq$\,0.01\,\Msun\ He-rich mantles
and $\simeq$\,10$^{-3.5}$\,\Msun\ H-rich envelopes.  
One such evolutionary model is run for
each 0.5$\sigma$ step in the $^{12}$C($\alpha,\gamma$)$^{16}$O
reaction rate probability distribution function.  We refer to this set of models as the
evolutionary DAVs. Each model used $\simeq$\,20,000 cells with a 30 isotope
nuclear reaction network and time resolution settings that consumed $\simeq$\,3 months of wall-clock time on 16 cores 
to complete $\simeq$ 550,000 time steps. 
Our models are similar to 
the lower resolution models used in \citet{timmes_2018_ab} and \citet{chidester_2021_aa}. 

We also use \code{wd\_builder} with \MESA\ version r15140 to build ab~initio WD models.
By ab~initio we mean calculations that begin with a hot WD model and an
assumed chemical stratification, as opposed to a hot WD model that is the
result of a stellar evolution calculation.  The imposed initial
\hydrogen, \helium, \carbon, \nitrogen, \oxygen, \neon[22], and \iron\ 
mass fraction profiles are taken from the evolutionary DAV models after 
the first thermal pulse on the asymptotic giant branch, defined by the first time the luminosity $L>10^4 L_\odot$. 
The H envelope is then thinned to mimic the H envelope thickness of the evolutionary DAVs,
the mass fraction profiles are smoothed at chemical transitions,
and mass location where \oxygen\ and \carbon\ exchange dominance is
taken to the average mass location for all $\sigma_i$.
This initial conditions procedure is done for DAV and DBV \code{wd\_builder} models.  
The DBV \code{wd\_builder} models are then stripped of their H envelope.

We use ab~initio WD models because 
they allow a more rapid exploration of the different WD classes, CO WD masses, and envelope masses
needed to preliminary assess the robustness of our results.  A potential disadvantage, 
or advantage, of ab initio WD models is that the imposed initial chemical stratification
may not be attainable by a stellar model evolved from the ZAMS.  

We use \GYRE\ release 6.0 \citep{townsend_2013_aa,townsend_2018_aa} embedded in \MESA\ version r15140
to calculate the adiabatic pulsation properties as a WD model evolves.
All the \MESA\ + \GYRE\ models begin from the top of the WD cooling track, terminate as cool WDs, 
and include the effects of element diffusion.
Details of the \MESA\ and \code{wd\_builder} models are 
in the files to reproduce our results at 
doi:\dataset[10.5281/zenodo.6779983]{[https://doi.org/10.5281/zenodo.6779983}.

We end this section by pointing out that we are not advocating for any specific model,
or any specific setting used by an evolutionary model (e.g., convective mixing parameters).
Rather our goal is to find, if there exists, potential signatures 
of the current experimental $^{12}$C$(\alpha,\gamma)^{16}$O reaction rate 
probability distribution function
in the pulsation periods of variable WD models. 
If such signatures exist and appear across the model space (including other researcher's variable WD models), 
then future uncertainty quantification studies could explore the impact of specific settings.

\begin{figure}[!ht]
    \centering
    \includegraphics[trim = {0.0cm 0.0cm 0.0cm 0.0cm },clip,width=\columnwidth]{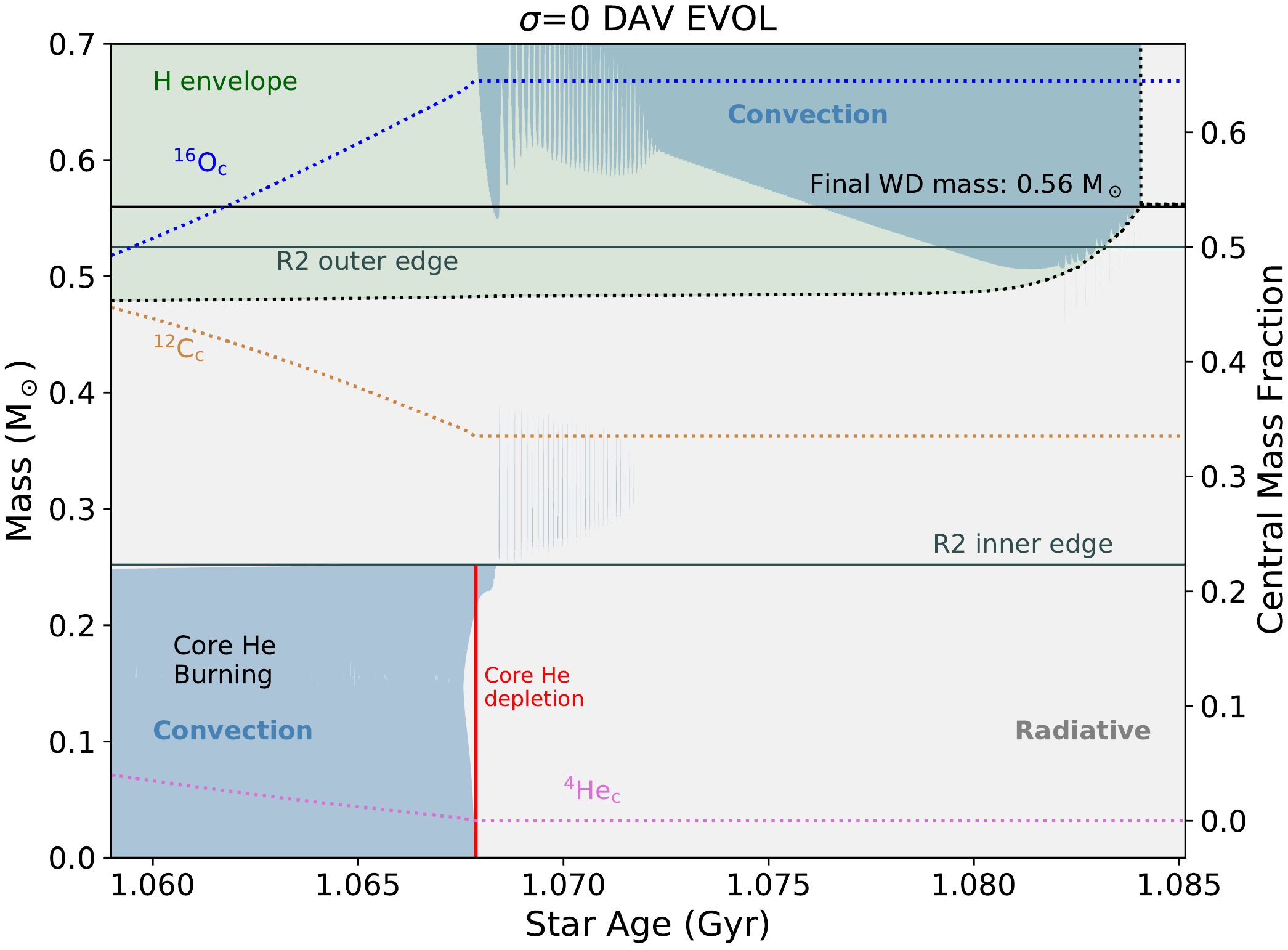}
    \caption{Kippenhahn diagram near core He depletion for the $\sigma$\,=\,0 evolutionary DAV model. Blue shading represents convective regions, gray the radiative regions, green the H envelope, and the red line marks core He depletion. Labeled are the eventual WD mass and the edges of radiative He burning region R2. Also shown is the evolution of the central \helium, \carbon, and \oxygen\ mass fractions.}
    \label{fig:convection_diagram}
\end{figure}

\section{Results}\label{sec:results}

\subsection{Evolutionary DAVs}\label{sec:detail}

Figure \ref{fig:rate_ratios} shows the mass fraction profiles of the evolutionary DAV models
for the 13 $\sigma_i$ reaction rates. We describe the labeled regions and chemical transitions.

Region R1 extends from the center to the transition between
\oxygen\ and \carbon\ in the core, henceforth the O$\rightarrow$C
transition, encompassing the innermost $\simeq$\,0.3\,\Msun.  
The reaction rate uncertainties have a large impact in R1, with the
central \oxygen\ mass fraction ranging from 0.45 for $\sigma$\,=\,-3.0
to 0.77 for $\sigma$\,=\,3.0 in a regular pattern.  The \oxygen\ and
\carbon\ mass fraction profiles are flat because this region 
forms during convective core He burning (see discussion of
Figure~\ref{fig:convection_diagram}).  Deconvolving how much
\oxygen\ and \carbon\ is due to the $^{12}$C$(\alpha,\gamma)^{16}$O
reaction rate and how much is due to convective mixing processes,
which has its own uncertainties, appears challenging \citep{straniero_2003_aa, de-geronimo_2017_aa}.

Region R2 extends from the O$\rightarrow$C transition to the
transition between \carbon\ and \helium, henceforth the
C$\rightarrow$He transition, encompassing $\simeq$\,0.25\,\Msun.  The
reaction rate uncertainties also have a large impact in R2, with the
\carbon\ and \oxygen\ mass fraction profiles showing a regular pattern
with $\sigma_i$.  This carbon-rich region forms during thick-shell He
burning under radiative conditions (see Figure~\ref{fig:convection_diagram}), 
mitigating the impact of convective mixing processes.  The 
significant spread in the \oxygen\ and \carbon\ mass fraction profiles is set
by the $^{12}$C$(\alpha,\gamma)^{16}$O reaction rate probability distribution function.  
The broad trend of the
\carbon\ mass fraction increasing and the \oxygen\ mass fraction
decreasing with enclosed mass reflects the decreasing temperature and density
in the radial direction during the formation of R2.

Region R3 extends from the C$\rightarrow$He transition to the
He$\rightarrow$H transition.  In the He-rich mantle of R3, the more
irregular pattern of the \oxygen, \carbon, and \helium\ mass fraction
profiles reflect variations in the thermal pulse histories.  Region R4
extends from the He$\rightarrow$H transition to the surface. The
$^{12}$C$(\alpha,\gamma)^{16}$O reaction rate does not have a strong
role in R4. In addition, the four most
abundant trace isotopes \neon[22], \neon[20], \nitrogen, and \iron\ do
not show a strong dependence on $\sigma_i$ in any region.

Figure~\ref{fig:convection_diagram} shows the evolution of the
$\sigma$\,=\,0 model near core He depletion and the
onset of thick-shell He burning. Region R1 in the final WD model is
formed during convective core He burning.
The extent of the convective core does not intrude into region R2.
Core convection ceases at core He
depletion, marked by the red line. The O-rich R1 region then becomes 
radiative for the remainder of the evolution.    Region R2 in the final WD model 
is formed after core He depletion during radiative thick-shell He burning.
The resulting C-rich R2 region then remains radiative for the remainder of the evolution. 

\begin{figure}
    \centering
    \includegraphics[trim = {0.0cm 0.0cm 0.0cm 0.0cm },clip,width=\columnwidth]{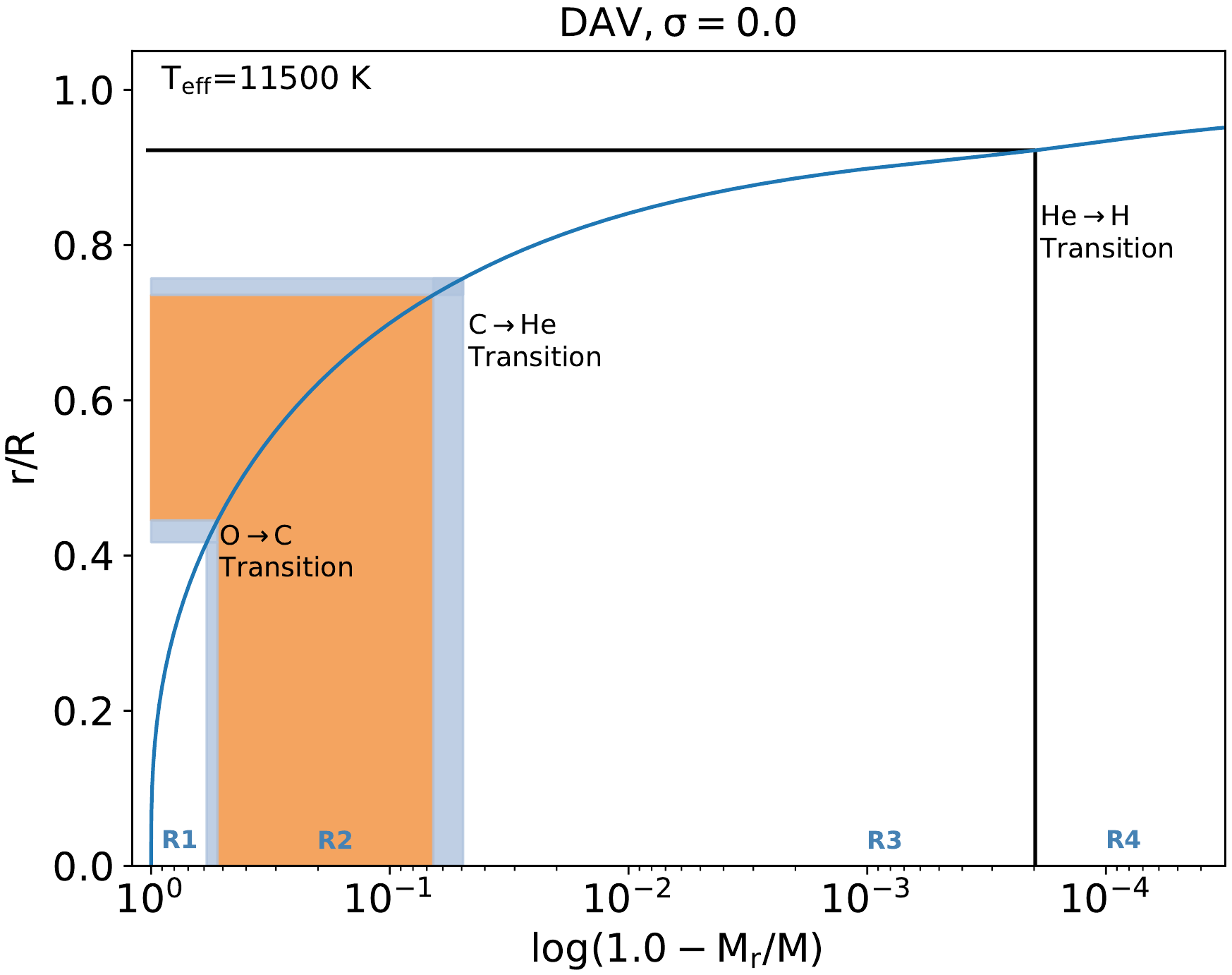}
    \caption{Mass-radius relation with key regions and chemical transitions annotated for the $\sigma$\,=\,0 evolutionary DAV model when it has cooled to \Teff\,=\,11,500~K. The shaded interiors highlight the regions that can most directly probe the $^{12}$C($\alpha,\gamma$) reaction rate.}
    \label{fig:mass-radius}
\end{figure}

It is useful to reference features with respect to mass or
radius.  Figure \ref{fig:mass-radius} thus shows the mass-radius
relation for the $\sigma$\,=\,0 model with regions and transitions
labeled. 

Some pulsation modes are more informative of these four regions than others.  
This can be due to a resonance, or near-resonance, between the mode's radial
wavelength and thickness of one or more of the composition layers
\citep{brassard_1991_aa}. As these modes traverse a
composition gradient within a local resonance region, they are
partially reflected and become ``trapped'' within the local layer \citep{winget_1981_aa}. Such modes 
are identified by showing a local minima in a kinetic energy diagram. 
We choose to refer to any mode displaying a local minima in the kinetic energy diagram 
as a trapped mode, regardless of the resonant region's location (e.g., upper or deeper layers).
Modes trapped by the upper layers can reveal insights about WD envelopes
\citep{kawaler_1990_aa, brassard_1992_ab, kawaler_1995_aa, costa_2008_aa}.
Modes trapped deeper in the WD can be sensitive to the different regions, 
and thus reveal insights on the interior chemical profiles \citep{brassard_1992_aa, corsico_2002_aa, giammichele_2017_ab}.
In particular, g-modes trapped by R2 can probe the current experimental
$^{12}$C($\alpha,\gamma$) reaction rate probability distribution function.

The \BV\ frequency $N$ is a characteristic frequency for pulsations, specifically
the frequency of oscillation about an equilibrium position under gravity:

\begin{equation}
N^2 = \frac{g^2 \rho}{P} \frac{\chiT}{\chir} (\grada - \gradT + B) \ ,
\label{eq:bv}
\end{equation}

\noindent
where
$g$ is the gravitational acceleration,
$\rho$ is the mass density,
$P$ is the pressure,
$T$ is the temperature,
$\chiT$ is the temperature exponent $\partial({\rm ln}P)/\partial({\rm ln}\rho)|_{T,\mu_I}$,
$\chir$ is the density exponent $\partial({\rm ln}P)/\partial({\rm ln}T)|_{\rho,\mu_I}$,
$\grada$ is the adiabatic temperature gradient,
$\gradT$ is the actual temperature gradient, and
$B$ is the Ledoux term that accounts for composition gradients \citep[e.g.,][]{hansen_1994sipp.book.....H,fontaine_2008_aa}. 
The implementation of Equation~\ref{eq:bv} in \MESA\ is described in \citet{paxton_2013_aa}. 

\begin{figure}[!ht]
    \centering
    \includegraphics[trim = {0.0cm 0.0cm 0.0cm 0.0cm },clip,width=\columnwidth]{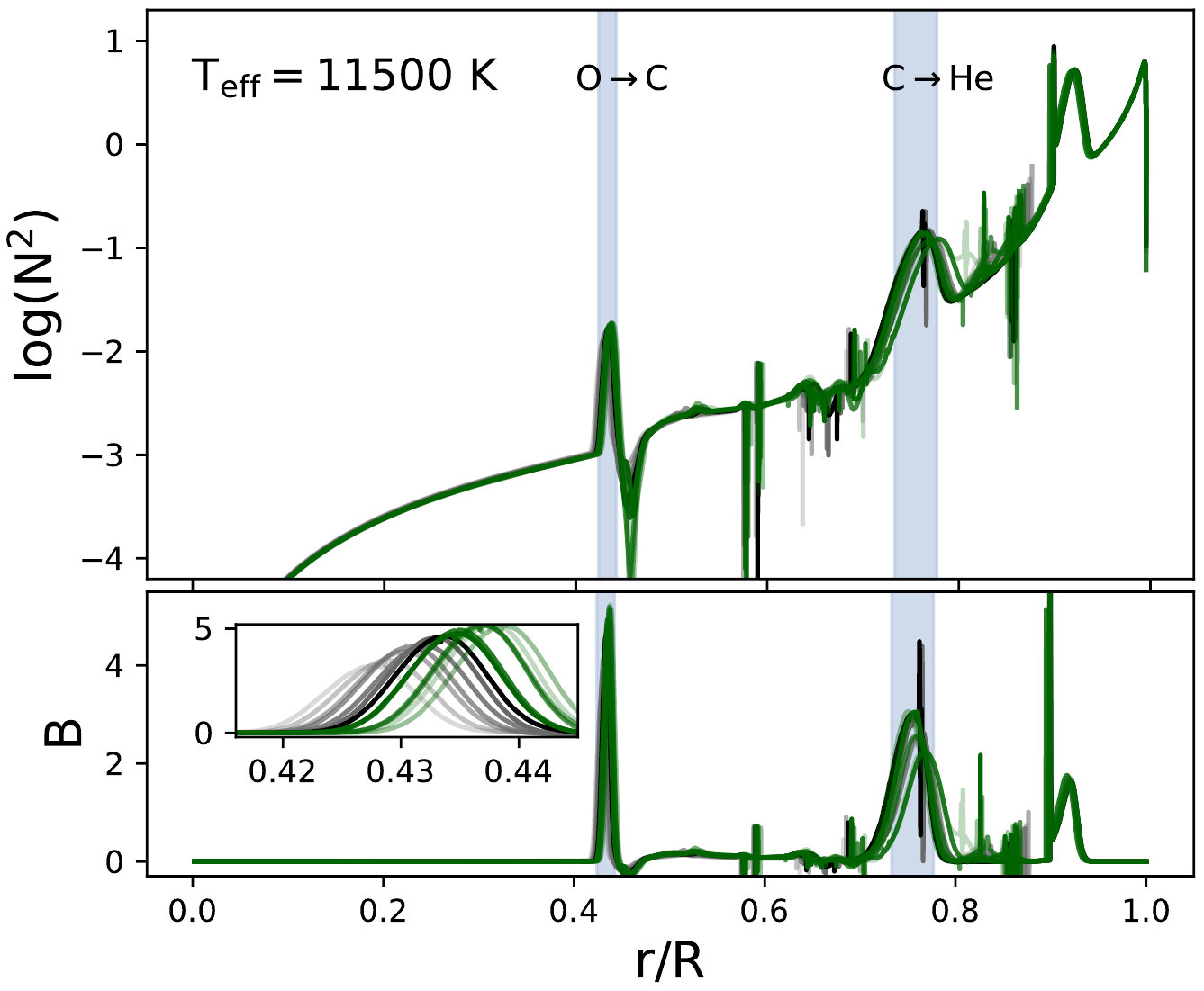}
    \caption{\textit{Top:} \BV\ frequency profiles as a function of fractional radius for the evolutionary DAVs after they have cooled to \Teff=11,500\,K.  As in Figure~\ref{fig:rate_ratios} the nominal $\sigma$\,=\,0 reaction rate is the black curve, positive $\sigma_i$ are green curves, and negative $\sigma_i$ are gray curves. Blue bands mark the O$\rightarrow$C and C$\rightarrow$He transition regions for all $\sigma_i$. \textit{Bottom:} The Ledoux $B$ term in Equation~\ref{eq:bv} as a function of fractional radius. An enlarged view of the first peak in $B$ at the O$\rightarrow$C transition is shown in the inset plot.}
    \label{fig:N2_bruntB_dav}
\end{figure}

Figure~\ref{fig:N2_bruntB_dav} shows the \BV\ frequency and the Ledoux $B$ term profiles
as a function of fractional radius after the 13 $\sigma_i$ evolutionary DAVs have cooled to \Teff\,=\,11,500\,K.  
The composition gradients in the O$\rightarrow$C and C$\rightarrow$He transition regions, highlighted by the blue bands, 
induce bumps in the Ledoux B term profile and thus bumps in the $N^2$ profile.
The first peak is located at the O$\rightarrow$C transition and magnified in the inset plot.
As $\sigma_i$ increases from -3.0 to 3.0, the location and magnitude of $B$ increase with radius in a near-regular pattern.

The kinetic energy $E_{\rm kin}$ of each g-mode can be expressed as \citep{unno_1989_aa, corsico_2002_aa}

\begin{equation}
    E_{\rm kin}=\frac{GMR^2}{2} \omega_n^2 \int_0^1x^2 \rho \left[ x^2 y_1^2 + x^2 \frac{\ell(\ell+1)}{(C_1 \omega_n^2)^2}y_2^2 \right] dx \ , 
    \label{eqn:ekin}
\end{equation}
where 
$n$ is the radial order,
$\ell$ the spherical harmonic degree,
$G$ the gravitational constant, 
$M$ the stellar mass, 
$R$ the stellar radius, 
$\omega_n^2$\,=\,$f_n^2 (GM/R^3)^{-1}$ the dimensionless eigenfrequency,
$f_n = 2\pi/P_n$ the frequency,
$P_n$ the period,
$C_1$\,=\,$x^3 (M/M_r)$ the scaled density, 
$r$ the radial distance from the center,
$M_r$ the stellar mass enclosed at radius $r$,
$x$\,=\,$r/R$ the scaled radius, 
and 
$y_1$ and $y_2$ are the dimensionless \citet{dziembowski_1971_aa} eigenfunctions.

\begin{figure}
    \centering
     \includegraphics[trim = {0.0cm 0.0cm 0.0cm 0.0cm},clip,width=\columnwidth]{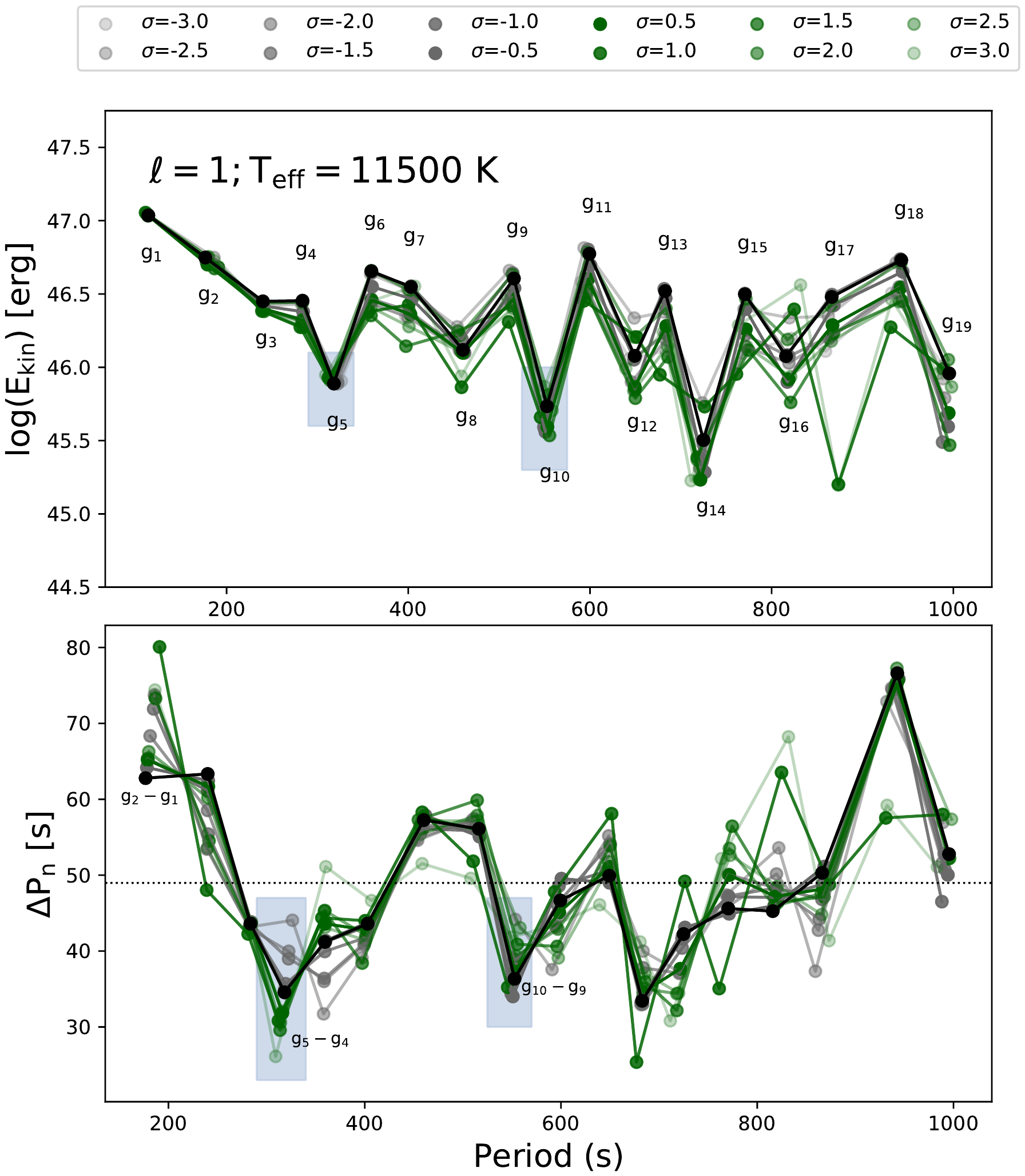}
    \caption{\textit{Top:} Kinetic energy of adiabatic g-modes for radial orders $n$\,=\,1--19 and harmonic degree $\ell$\,=\,1 after the evolutionary DAVs have cooled to \Teff\,=\,11,500\,K.  The colors are as in Figure~\ref{fig:rate_ratios}. Local minima identify trapped modes. The $g_5$ and $g_{10}$ trapped modes are highlighted with blue bands. \textit{Bottom:} Period spacing diagram with the local minima at $g_5$ and $g_{10}$ highlighted by blue bands. The mean period spacing for $\sigma$\,=\,0 model is marked by the dashed line.}
    \label{fig:trapped_modes}
\end{figure}

We numerically integrate Equation~\ref{eqn:ekin} for g-modes of radial orders $n$\,=\,1--19 and harmonic degree $\ell$\,=\,1.  
Unless otherwise specified, all g-modes considered have $\ell$\,=\,1.
Figure~\ref{fig:trapped_modes} shows the resulting $E_{\rm kin}$ and period spacing $\Delta P_n$\,=\,$P_{n+1}-P_n$ diagrams for the
evolutionary DAVs when they have cooled to \Teff\,=\,11,500\,K. 
Trapped modes are identified by minima in $E_{\rm kin}$
and by a corresponding minima in the period spacing \citep[e.g., ][]{winget_1981_aa, brassard_1991_aa}.
Figure~\ref{fig:trapped_modes} highlights two strong local minima, the $g_5$ and $g_{10}$ trapped modes.

The frequency of an adiabatic g-mode is the area under a curve known as 
a weight function ${\rm d}\zeta/{\rm d}r$

\begin{equation}
f^2 =  \zeta = \int_{r=0}^{r=R} \frac{{\rm d}\zeta}{{\rm d}r} \cdot {\rm d}r
\ ,
\label{eq:weight_function}
\end{equation}

\noindent
where, following \citet{kawaler_1985_aa}, 

\begin{equation}
  \frac{{\rm d}\zeta}{{\rm d}r} = \frac{[C({\bf y},r) + N({\bf y},r) + G({\bf y},r)] \rho r^2 }{\int_{r=0}^{r=R} T({\bf y},r) \rho r^2 {\rm d}r}
\ 
\label{eq:dzdr}
\end{equation}

\noindent
and 
$C({\bf y},r)$ varies with the Lamb frequency,
$N({\bf y},r)$ contains the \BV\ frequency,
$G({\bf y},r)$ involves the gravitational eigenfunctions,
$T({\bf y},r)$ is proportional to the kinetic energy density,
and ${\bf y} = (y_1,y_2,y_3,y_4)$ are the \citet{dziembowski_1971_aa} variables.

\begin{figure}
    \centering
    \includegraphics[trim = {0.0cm 0.0cm 0.0cm 0.0cm },clip,width=\columnwidth]{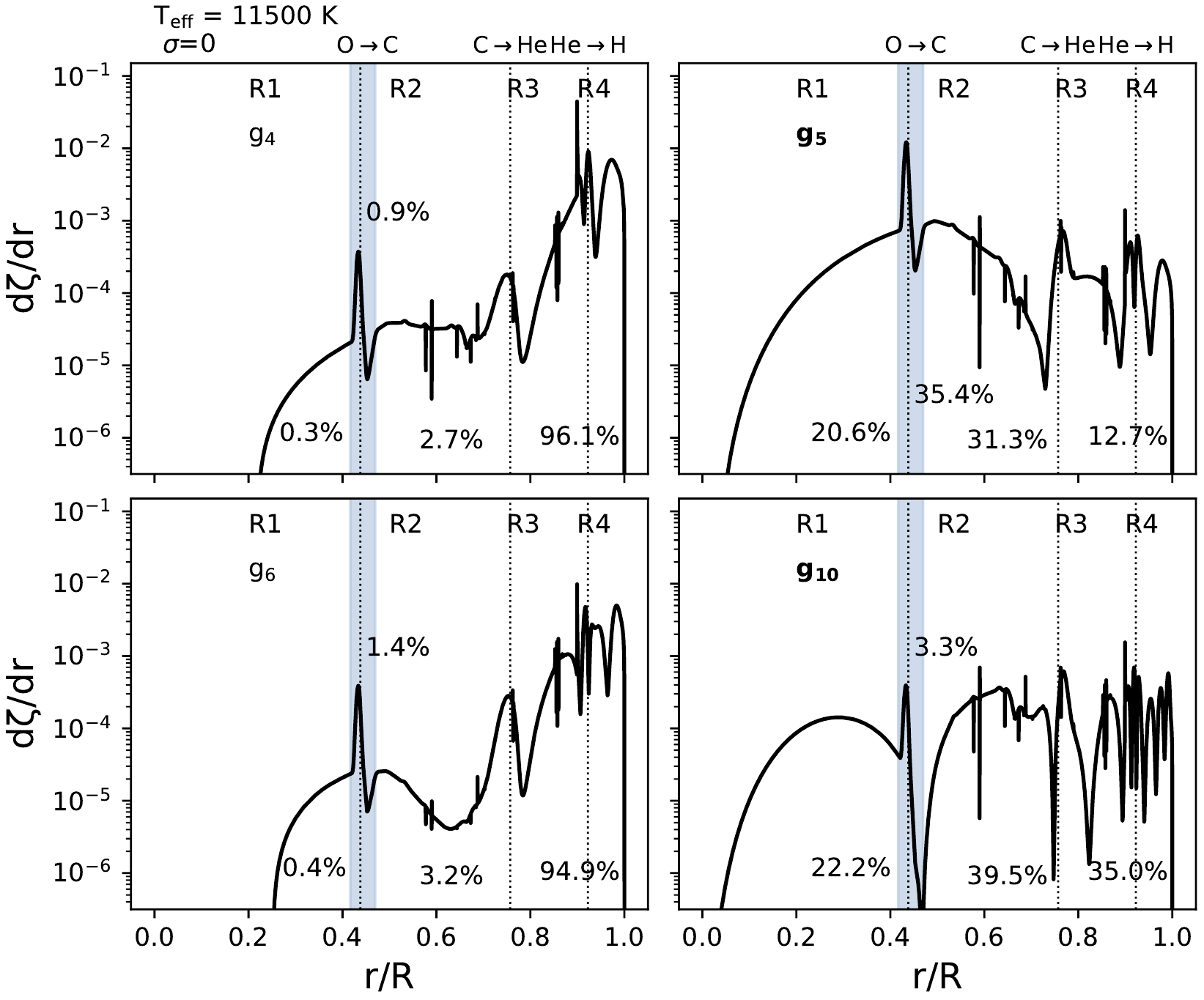}
    \caption{Weight functions of the $g_4$, $g_5$, $g_6$, and $g_{10}$ modes for the $\sigma$\,=\,0 model after it cools to \Teff\,=\,11,500\,K. Trapped modes are in bold font.     Weight functions returned by \code{GYRE} are normalized to have unit area under their curve. The fractional area under the curve of region R1, the O$\rightarrow$C transition, region R2, and the sum of R3 and R4 are given.}
    \label{fig:weight}
\end{figure}

Figure~\ref{fig:weight} shows the weight functions 
of the $g_4$, $g_5$ $g_6$, and $g_{10}$ modes.
The $g_5$ and $g_{10}$ trapped modes have larger weight functions in
regions R1 and R2 compared to the $g_4$ and $g_{6}$ non-trapped modes.
The frequency of the trapped modes is thus more strongly weighted by regions R1 and R2.
The  $g_{10}$ weight function is more equally distributed 
than the $g_5$ weight function, and the peak at the O$\rightarrow$C 
transition is about the same height as the other peaks. Both factors decrease
the ability of $g_{10}$ to probe the O$\rightarrow$C transition and region R2.  
In contrast, the peak in the $g_5$ weight function at the O$\rightarrow$C transition 
is the largest peak. About 35\% of the $g_{5}$ frequency comes from 
the O$\rightarrow$C transition, and $\simeq$\,67\% from 
the O$\rightarrow$C transition and R2.

Taken together, these weight functions indicate that R2 contributes $\simeq$\,30--40\% 
and R1 contributes $\simeq$\,20\% to the periods of $g_5$ and $g_{10}$.
This suggests that applying our results,
conducted over the current experimental $\pm$\,3$\sigma$ uncertainty in the $^{12}$C$(\alpha,\gamma)^{16}$O 
reaction rate probability distribution function,
to the derived pulsation spectrum of observed variable WD may face challenges obtaining
precision constraints on the reaction rate.

\begin{figure}
    \centering
    \includegraphics[trim = {0.1cm 1.cm 0.6cm 2cm },clip,width=0.99\columnwidth]{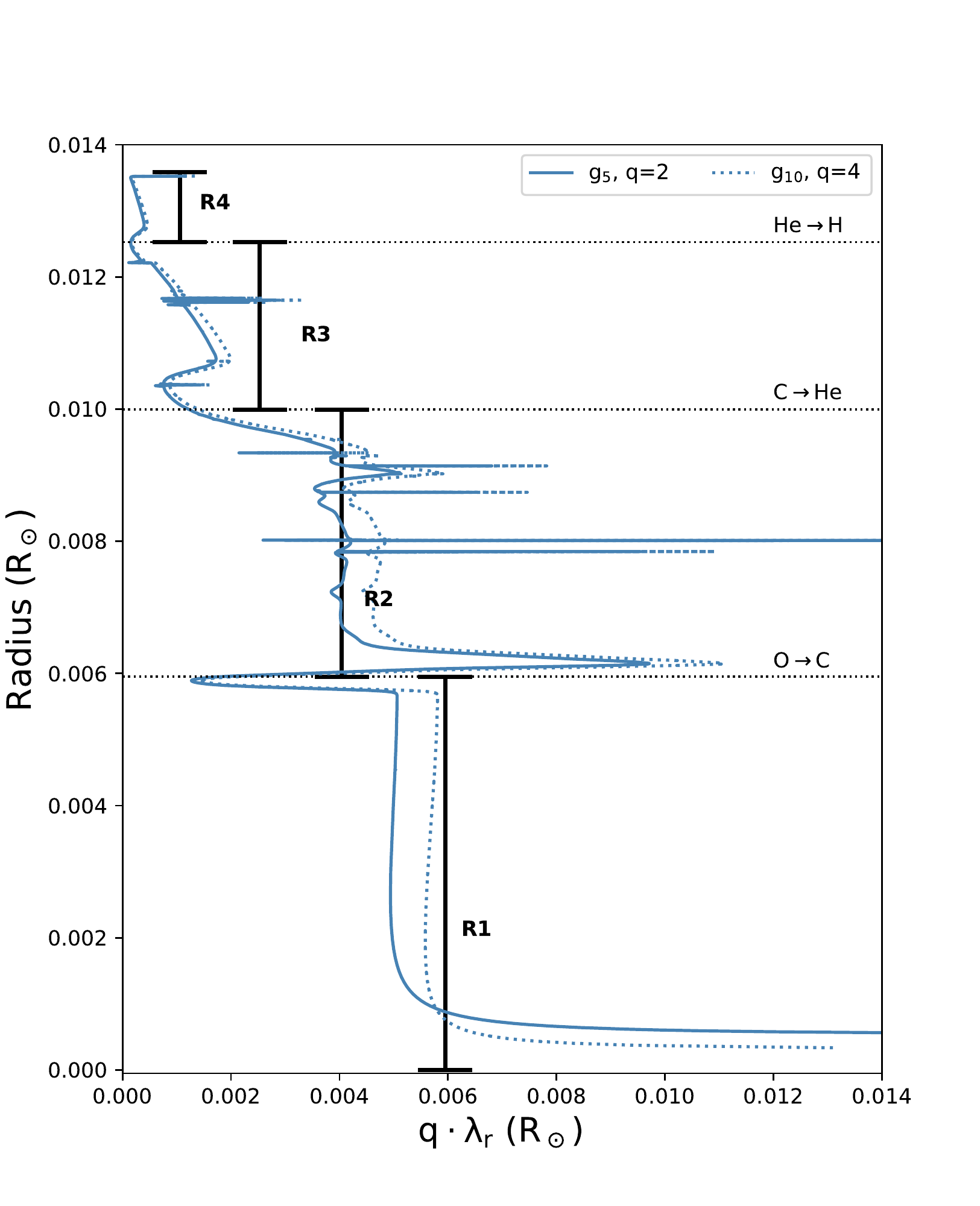}
    \caption{Integer multiples $q$ of the radial wavelength $\lambda_r$ profiles versus radius for the $g_5$, and $g_{10}$ modes, for $\sigma_0$. The trapped $g_5$ mode is shown by the dark blue solid line, and the dotted curves depict the trapped $g_{10}$ mode.  Solid black segments depict the width of the regions R1, R2, R3, and R4, as defined by distance between labeled chemical transitions.}
    \label{fig:lambda}
\end{figure}

\begin{figure}[!ht]
    \centering
    \includegraphics[trim = {0.0cm 0.0cm 0.0cm 0.0cm },clip,width=\columnwidth]{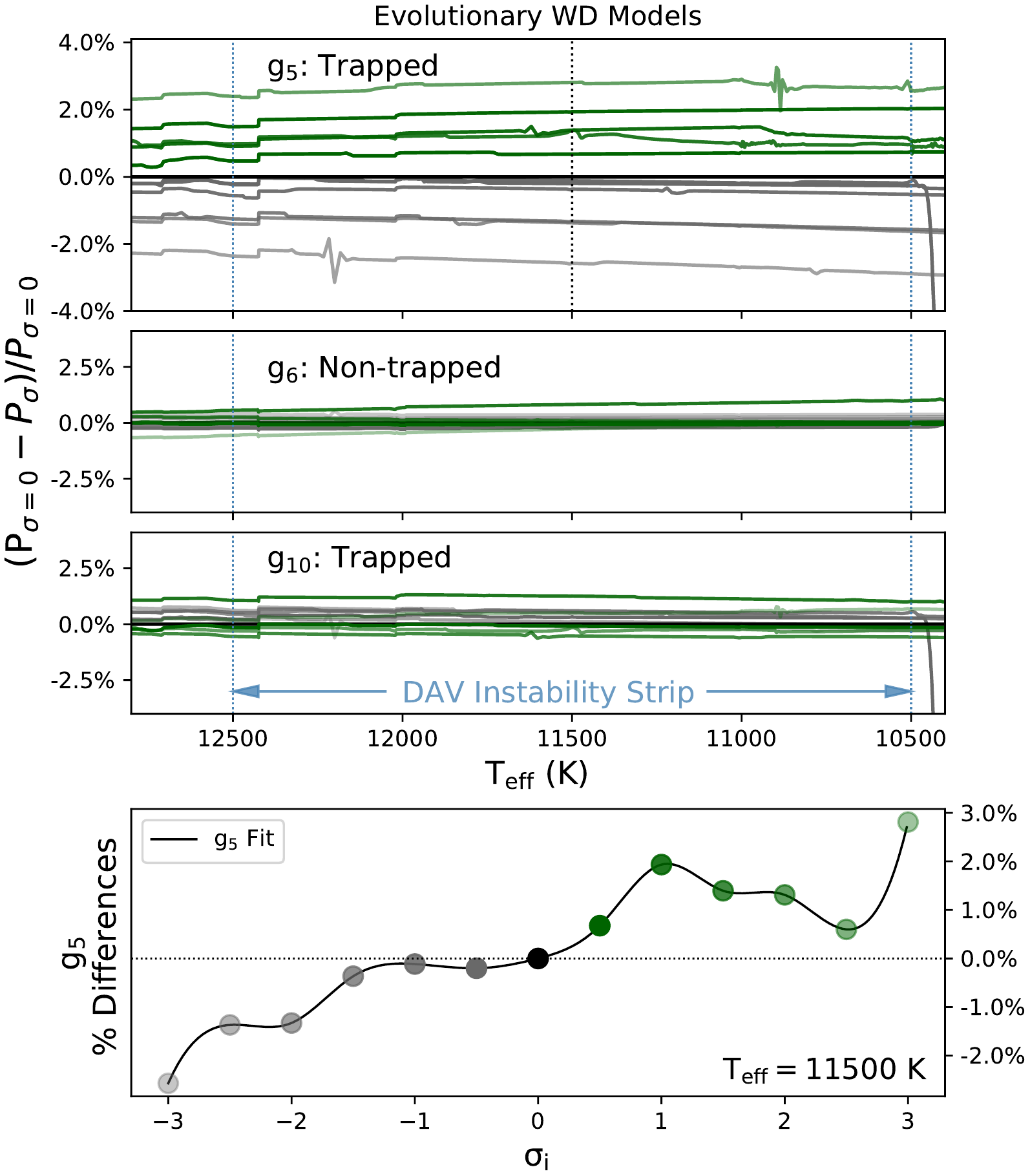}
    \caption{\textit{Top panels:} Relative period differences from $\sigma$\,=\,0 for the $g_5, g_6$, and $g_{10}$ modes as the evolutionary DAVs cool.  Colors for the $\sigma_i$ follow Figure \ref{fig:rate_ratios}.  The range of \Teff\ in observed DAV WDs is marked, with the vertical dashed black line selecting the \Teff\,=\,11,500~K midpoint. \textit{Bottom panel:} Relative period differences for $g_5$ versus the $^{12}$C($\alpha$,$\gamma$)$^{16}$O reaction rate uncertainties $\sigma_i$ at \Teff\,=\,11,500~K.  Scatter points are the raw data values and the curve is a polynomial fit.}
    \label{fig:dPs_dav}
\end{figure}

Resonance regions can also be identified by checking when the 
width of a chemical stratification region is equal to an integer 
number of radial wavelengths, $\lambda_r = 2\pi / k_r$ where the 
wavenumber is

\begin{equation}
    k_r^2=\frac{\ell(\ell+1)}{r^2f^2S_\ell^2}(f^2-N^2)(f^2-S_l^2)
\end{equation}

\noindent
and $S_\ell$ is the Lamb frequency.  
Figure~\ref{fig:lambda} shows integer multiples $q$ of $\lambda_r$ 
for the $g_5$ and $g_{10}$ modes versus radius for $\sigma_0$. 
When $q$\,$\cdot$\,$\lambda_r$ is near a black segment, the g-mode resonates, 
or nearly resonates, with the region's width.
The 2$\cdot \lambda_{r}$ curve for $g_5$ lies close to the width
of R2, identifying R2 as the resonant cavity for the $g_5$ mode.
For the $g_{10}$ mode the 4$\cdot \lambda_r$ curve is close to the widths of
R1 and R2.  We also verified 6$\cdot \lambda_{r}$ is close to the width of R3. 
Larger $q$ values for $g_{10}$ may resonate with R4.  
As $g_{10}$ resonates with multiple regions, the mode is not uniquely trapped.
This is commensurate with the more uniform distribution of peaks 
in the $g_{10}$ weight function of Figure~\ref{fig:weight}.

Figure~\ref{fig:dPs_dav} shows the history of the relative period
differences for $g_5$, $g_6$, and $g_{10}$ as the evolutionary DAVs cool.  
The $g_5$ trapped mode shows the most distinctive trend in the
period with $\sigma_i$ out of every g-mode in the range $1\leq n \leq 20$.    
The relative period differences
span $\simeq$\,$\pm$2\%, with positive $\sigma_i$ yielding shorter periods
and negative $\sigma_i$ yielding longer periods.  The $g_{10}$ trapped
mode is not as distinctive, showing a smaller spread in periods with
$\sigma_i$. The $g_6$ non-trapped mode shows little spread with
$\sigma_i$ and no distinctive trend with period.  
While Figure~\ref{fig:dPs_dav} highlights $g_6$ and $g_{10}$, 
we emphasize that no other g-mode within $1\leq n \leq 20$ shows any distinctive period pattern with $\sigma_i$.

Figure~\ref{fig:dPs_dav} also shows the relative period differences
for $g_5$ with respect to $\sigma_i$ at \Teff\,=\,11,500~K.  The
relative period differences increase from $\simeq$\,$-$2.0\% for
$\sigma$=$-$3.0 to $\simeq$\,3.0\% for $\sigma$\,=\,3.0.  The
relationship is non-monotonic due to variations in the location of the
C$\rightarrow$He transition, which impacts the width of R2 where the
$g_5$ trapped mode resonates.  Minimizing variations in this
location may increase the monotonicity of this relationship.

\begin{figure*}[!ht]
    \centering
    \includegraphics[width=0.99\doublewide]{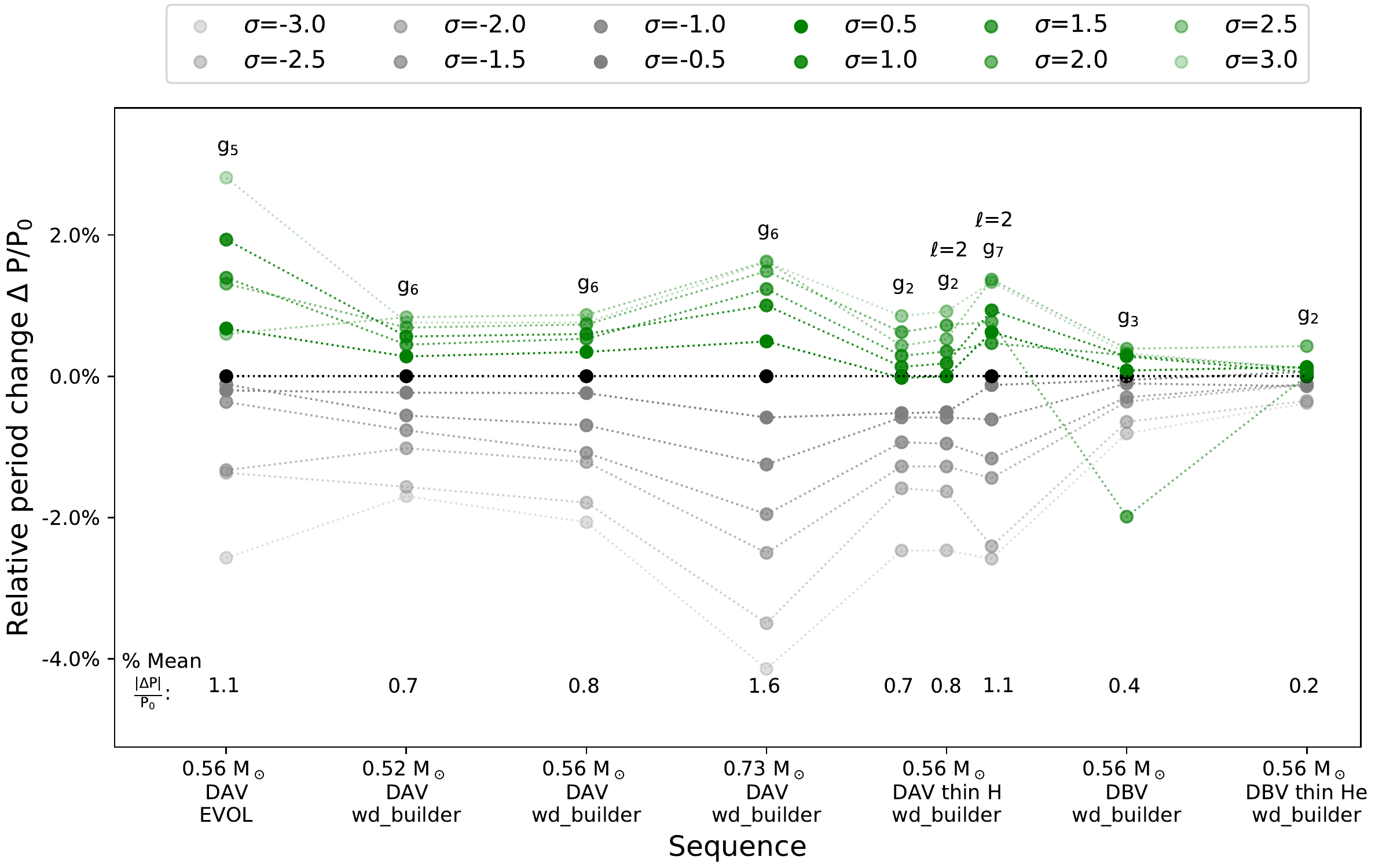}
    \caption{Relative period differences for model sequences of varying WD masses, shell masses, and classes.  The \code{wd\_builder} DAVs were measured at \Teff\,=\,11,500~K for the \code{wd\_builder} DBVs were measured at \Teff\,=\,25,000~K. Labeled are the $\pm$\,3\,$\sigma$ uncertainties on the experimental $^{12}$C$(\alpha,\gamma)^{16}$O reaction rate in steps of 0.5$\sigma$, the trapped g-mode that most distinctly probes the radiatively-formed, carbon-rich region R2, and the mean relative period difference for each sequence.}
    \label{fig:models}
\end{figure*}

\subsection{Variations}\label{sec:all4}

We have presented evidence that adiabatic g-modes trapped by the
radiatively formed carbon-rich layer in CO WD models offer potentially
useful probes of the current experimental $^{12}$C$(\alpha,\gamma)^{16}$O reaction 
rate probability distribution function. In this section we give a preliminary assessment of the
robustness of this result by sampling different WD classes, masses,
and shell masses.  Each sequence contains 13 $\sigma_i$ models.  
For each sequence, we verify the existence of a trapped or partially trapped mode that probes region R2 
as indicated by the sequences's kinetic energy diagram and weight functions. We also confirm that the 
sequences's R2 trapped g-mode gives the most distinct $\Delta P/P_0$ versus $\sigma_i$ relationship.   

Figure~\ref{fig:models} shows the results of this survey.  The x-axis
is the sequence  and the y-axis is relative period difference $\Delta P/P_0$ for 
the labeled R2 trapped g-mode. Proceeding from left to right,
the first sequence is the evolutionary DAVs which are analyzed in detail above.

The next three sequences explore the impact of the WD mass using
\code{wd\_builder} DAV models of 0.52, 0.56, and 0.73~\Msun, respectively.
All three sequences show $g_6$ is the most distinctive adiabatic
trapped g-mode that probes R2.  The differences between the
0.56~\Msun\ evolutionary DAVs and the 0.56~\Msun\ \code{wd\_builder} DAVs is
due to their different composition profiles (see
Section~\ref{sec:models}).  This suggests that the most distinctive
trapped g-mode depends on model details.  The 0.52 and 0.56\,\Msun\,
sequences are similar to one another, span a smaller $\Delta
P/P_0$\,$\simeq$\,$\substack{+1 \\ -2}$\% range than the evolutionary
DAVs, and show non-monotonic spacings with $\sigma_i$.  The 0.73 \Msun\,
sequence spans $\Delta P/P_0$\,$\simeq$\,$\substack{+2 \\ -4}$\% and
shows a monotonic spacing with $\sigma_i$.  This suggests that more
massive WDs may give stronger signals with this method.

The fifth sequence shows the impact of a H envelope that is 10 times thinner than the 0.56 \Msun\ \code{wd\_builder} DAVs.  
One dipole $\ell$\,=\,1 and two quadruple $\ell$\,=\,2 trapped g-modes are shown.
The $g_2$ trapped mode with $\ell$\,=\,1 shows a regular pattern with $\sigma_i$ over the range $\Delta P/P_0$\,$\simeq$\,$\substack{+1 \\ -2.5}$\%.
The $g_2$ trapped mode with $\ell$\,=\,2 is similar to its $\ell=1$ counterpart, and 
the $g_7$ trapped mode with $\ell$\,=\,2 shows a larger $\Delta P/P_0$\,$\simeq$\,$\substack{+1.5 \\ -2.5}$\% range.
This sequence suggests that harmonic degrees $\ell$\,=\,1 and/or $\ell$\,=\,2 can have R2 trapped modes that distinguish $\sigma_i$.

The next sequence shows the \code{wd\_builder} DBVs. A regular pattern with $\sigma_i$
for the $g_3$ trapped mode emerges, and the sequence spans $\Delta P/P_0$\,$\simeq$\,$\substack{+0.5 \\ -1}$\%.
The $\sigma$\,=\,2.0 point in this sequence is an outlier 
that we cannot explain.
The last sequence shows the impact of a He envelope that is 10 times thinner
than the 0.56 \Msun\ \code{wd\_builder} DBVs. A regular pattern with $\sigma_i$ persists
at the  $\Delta P/P_0$\,$\simeq$\,$\substack{+0.5 \\ -0.5}$\% level for the $g_2$ trapped mode.

For all the sequences, the positive $\sigma_i$ have shorter periods
than the $\sigma$\,=\,0 model.  Following \citet{chidester_2021_aa},
the main contributors to period differences are changes in the local
pressure scale height $H$, mean molecular weight $\mu_i$, temperature,
density exponent $\chi_{\rho}$, and first adiabatic index $\Gamma_1$.
Changes in these parameters compete with one another to drive the
overall period difference.  Models with larger mass fractions of
\neon[22] give shorter periods due to a smaller $H$.  As increasing
the \COrate\ increases the \oxygen\ content in R2, smaller $H$ drive
shorter periods.  Similar logic applies to negative $\sigma_i$ models
showing longer periods. 

\section{Sensitivities}\label{sec:sensitivities}

There are many potential sensitivities that we have not investigated.
We highlight four and the edges of R2 that they affect.

\subsection{Width of the O$\rightarrow$C transition}\label{sec:width}

Mode trapping by R2 may depend on the width of the O$\rightarrow$C transition. 
Increasing the O$\rightarrow$C transition width decreases the $\mu_i$ gradient,
which in turn may decrease the trapping ability of R2.
The width of the O$\rightarrow$C transition in our models is relatively narrow,
in agreement with the \MESA\ models of \cite{pepper_2022_aa}, 
while the O$\rightarrow$C transition in \cite{corsico_2002_aa} is considerably wider.  
\cite{corsico_2002_aa} also find that their adiabatic $g_5$ mode depends on the  O$\rightarrow$C transition and R2.  
However, their $g_5$ mode is a local maximum in their $E_{\rm kin}$ diagram, meaning it is not a trapped mode.  
With mode trapping suppressed for a wide enough O$\rightarrow$C transition, the kinetic energy from trapped modes in R2 is released.
As Equation~\ref{eqn:ekin} is weighted with the density, 
a non-trapped g-mode sensitive to R2 may appear as a local maximum in $E_{\rm kin}$, as found by \cite{corsico_2002_aa}.  
Relative extrema in $E_{\rm kin}$, not just minima, might find g-modes that probe R2 and thus also reveal inferences on $\sigma_i$.
Region R2 still exists, because it forms under thick-shell radiative He-burning conditions.
We caution that its trapping abilities, and thus our relative period shifts 
may change with the width of the O$\rightarrow$C transition.

\subsection{3$\alpha$ reaction rate probability distribution function}\label{sec:3a}

The 3$\alpha$ process, the fusion of three \helium\ nuclei into one
\carbon\ nucleus, impacts the innermost and outermost edges of R2.
The 3$\alpha$ process is followed by the subsequent
$\alpha$-capture reaction $^{12}$C$(\alpha,\gamma)^{16}$O.  The final
mass fractions of \carbon\ and \oxygen, under radiative burning
conditions, is determined by the competition between the 3$\alpha$ and
$^{12}$C$(\alpha,\gamma)^{16}$O reaction rates at a given temperature.
The feeding of \carbon, driven by the 3$\alpha$ process, occurs early
in the evolution when the mass fraction of \carbon\ is low and
\helium\ is high (the carbon bump at the outermost edge of R2).
Oxygen production occurs later by $\alpha$-capture on the freshly
produced \carbon\ (the O$\rightarrow$C transition at the innermost edge
of R2).  Current estimates of the uncertainty in the 3$\alpha$
reaction rate are $\simeq$\,30\% over the regions of typical
astrophysical interest \citep{kibedi_2020_aa}. However, at lower
temperatures ($\lesssim$ 0.1 GK), the uncertainty is likely much
larger, because other reaction mechanisms become significant
\citep{suno_2016_aa}.  
Region R2 again still exists, and we again caution that the period shifts 
we find from the \COrate\ probability distribution may change 
when a 3$\alpha$ reaction rate probability distribution function is considered.

\subsection{Mixing during core He-depletion}\label{sec:mixing}

Mixing in low- and intermediate-mass stars during core He-burning
is particularly challenging to model \citep{salaris_2017_aa,jermyn_2022_aa} and impacts
the innermost edge of R2.
The radiative gradient  profile within a He core convection region develops a 
local minimum at some point during its evolution \citep[e.g., see Figure 6 of][]{paxton_2018_aa}.
With further outward propagation of the convective boundary, 
or the action of overshooting, the mixing of fresh He into 
the core can lower the radiative gradient 
throughout the core to such an extent that it equals the adiabatic gradient 
at the local minimum of the radiative gradient.  
When this happens, the convective region interior to the
minimum becomes decoupled from the region exterior to the minimum: the
convective core splits \citep{eggleton_1972_aa}. 
In addition, even small amounts of He added to the convective core enhances 
the rate of energy production by the $^{12}$C$(\alpha,\gamma)^{16}$O reaction.
The resulting increase in the radiative gradient can lead to rapid growth 
in the convective He core boundary (a ``breathing pulse''). The enhanced nuclear burning also 
increases the central \oxygen\ mass fraction. 
A concensus on breathing pulses being physical or numerical has not yet been reached
\citep{caputo_1989_aa, cassisi_2003_aa, farmer_2016_aa, constantino_2017_aa, paxton_2019_aa}. 
Region R2 persists, and we caution that the absolute period shifts we find may change with 
different treatments of core He-depletion in evolutionary models.

\citet{de-geronimo_2017_aa} found that overshooting during the core He burning
leaves imprints on the \BV\ frequency that result in absolute period differences of 
$\simeq$\,2--5 s on average, relative to models with zero overshooting.
Models that included overshooting had a larger central \oxygen\ mass fraction and an extended R1. 
They found that these results are, on average, independent of the ZAMS mass. 
\citet{de-geronimo_2017_aa} concluded their \COrate\ reaction rate
uncertainties were less relevant than their uncertainties from overshooting.
The \citet{kunz_2002_aa} reaction rate adopted in \citet{de-geronimo_2017_aa} 
is different in shape over the relevant temperature range than the \citet{deboer_2017_aa} reaction rate; 
see Figure 29 in \citet{deboer_2017_aa}. A simple scaling of the \citet{kunz_2002_aa} reaction rate 
is not the same as adopting a modern reaction rate sourced from a probability distribution function
\citep{ mehta_2022_aa}.
We suspect the \citet{de-geronimo_2017_aa} result is partially due to using a \COrate\ reaction rate 
that is $\simeq$\,10\% larger than their base reaction rate, as the highest rate considered.
In contrast our \COrate\ rate probability distribution function  spans $\simeq$\,0.5--1.5 times
our nominal $\sigma$\,=\,0 reaction rate (see Figure~\ref{fig:rate_ratios}).  
Therefore, we find larger overall average period differences from the \COrate\ reaction rate.
Future uncertainty quantification studies could explore a potential coupling 
between simultaneous variations in overshooting and the adopted \COrate\ reaction rate.
It is possible that a strong coupling could alter the R2 trapped mode properties of our models.

\subsection{Number of Thermal Pulses}\label{sec:tp}

The thermal pulse phase of evolution impacts the outermost edge of R2.
For the case of a fixed number of thermal pulses,
\citet{de-geronimo_2017_aa} found period differences of 
$\simeq$\,5--10 s for their 0.548\,\Msun\ model and 
$\simeq$\,2–3 s for their 0.837\,\Msun\ model.  This effect is mainly due to the 
C$\rightarrow$He transition being less pronounced in their more massive WD models. 
The impact of the thermal pulses in our evolutionary models is shown in Figure~\ref{fig:rate_ratios}.
Each of our 13 $\sigma_i$ models experienced $\simeq$\,14 thermal pulses, with 
the onset of each thermal pulse defined by the photon luminosity exceeding 10$^4$ \Lsun.
In contrast, our \code{wd\_builder} models were inferred from the chemical profiles 
at the first thermal pulse, fixing the number of thermal pulses for those sequences.  
The sensitivity of our results to the number of thermal pulses
can thus be estimated by comparing the $g_5$ trapped mode periods of 
our 0.56\,\Msun\ evolutionary and \code{wd\_builder} models.

We find a standard deviation of $\simeq$\,2.3 for the thermal pulses
and a standard deviation of $\simeq$\,4.5 for the \COrate.
This suggests that variations from the number of thermal pulses in our models 
is smaller than the variations from the \COrate\ reaction rate probability distribution function.
In contrast, \citet{de-geronimo_2017_aa} found larger variations from thermal pulses than the \COrate\ reaction rate. 
We speculate this difference is again due to our larger span of \COrate\ reaction rates.
Our $\simeq$\,14:1 thermal pulse ratio is also larger than the 10:3 thermal pulse ratio of \citet{de-geronimo_2017_aa},
which may strengthen our result that variations from the \COrate\ reaction rate probability distribution function
have a larger impact than the number of thermal pulses.
In addition, \cite{pepper_2022_aa} found the number of thermal pulses is dependent on the \COrate\ reaction rate,
which Figure~\ref{fig:rate_ratios} confirms. Also in agreement with \cite{pepper_2022_aa}, 
we find that smaller $\sigma_i$ increases the number of thermal pulses as smaller
reaction rates have larger He-shell masses at the onset of each pulse.

\section{Summary}\label{sec:summary}

We conducted a new search for signatures of the current experimental
$^{12}$C$(\alpha,\gamma)^{16}$O reaction rate probability distribution function
in the pulsation periods of CO WD models.
We found that adiabatic g-modes trapped or partially trapped by the interior
C-rich layer (region R2 in Figure~\ref{fig:rate_ratios}) offer the most direct probe
of this reaction rate because this region forms under radiative He burning conditions, 
mitigating the impact of uncertainties from convective mixing processes.  
We found an average spread in relative period shifts of $\Delta P/P \simeq \pm$\,2\% for the identified trapped
g-modes over the experimental $\pm$\,3$\sigma$ probability distribution function of 
the $^{12}$C$(\alpha,\gamma)^{16}$O reaction rate. 
We found the effect persists across the observed \Teff\ window
of DAV and DBV WDs, and for different WD masses and smaller H/He shell masses.
Figures \ref{fig:rate_ratios}, \ref{fig:dPs_dav}, and \ref{fig:models}
make the first direct quantitative connection between the pulsation periods of variable WD models 
and the current, experimental $^{12}$C$(\alpha,\gamma)^{16}$O  reaction rate probability distribution function. 

The C-rich layer is a ``sweet spot'' in DAV and DBV WD models for probing the $^{12}$C$(\alpha,\gamma)^{16}$O
reaction rate probability distribution function. 
Figure~\ref{fig:models} suggests a corresponding ``sweet spot'' of g-modes
with radial orders 2\,$\lesssim$\,$n$\,$\lesssim$\,7  can investigate R2.
This suggestion is complemented by an analysis from \cite{corsico_2002_aa}, who found 
that all g-mode periods $\gtrsim$\,500--600 s were trapped (or nearly trapped) in the
H-rich envelope.  They found the weight functions of those modes were
low in amplitude, and similar to one another, and concluded that mode
trapping vanishes for long periods (higher radial orders).  Thus,
inferences from trapped modes such as R2 are limited to
g-modes with periods $\lesssim$\,500--600\,s.  
This confines our models to have g-modes  $n$\,$\lesssim$\,10.

In every model sequence explored, the g-mode that best distinguishes the \COrate\ reaction rate follows five
specific patterns:

\begin{enumerate}
    \item The g-mode is trapped, as confirmed by its local minimum in $E_{\rm kin}$.
    \item  The g-mode best resonates with the R2 region; it has a $q \cdot \lambda_r$ matching best with the R2 region width, and gives a weight function substantially weighted in the R2 region. 
    \item The g-mode is within a radial order ``sweet spot" of $2 \lesssim n \lesssim 7$.
    \item The g-mode period is shorter for positive $\sigma_i$, and longer for negative $\sigma_i$. 
    \item The g-mode shifts are within the detectable range.
\end{enumerate}

The signatures persist because R2 forms under radiative helium burning conditions,
but could be sensitive to the couplings with other uncertainties (see Section \ref{sec:sensitivities}).
Out of the entire g-mode
spectrum, only one, the R2 trapped g-mode, consistently showed an
identifiable trace to the \COrate\ reaction rate probability distribution function.  Moreover, the R2
trapped mode consistently followed the 5 patterns listed above,
irrespective of model type, WD class, WD mass, and envelope thickness. 
Thus, the trapped R2 g-mode signature found is
the most direct information at tracing the \COrate\ reaction rate probablity distribution function using WD seismology.

The g-mode periods of observed variable WD are derived from a Fourier
analysis of the photometric light curves and are typically given to
6$-$7 significant figures of precision \citep[e.g.,][]{duan_2021_aa}.
Usually WD composition profile templates are fit to the extracted
g-mode period spectrum and other observed constraints (e.g., $\Teff$,
$\log g$) of a specific WD.  The root-mean-square residuals to the
$\simeq$\,150$-$400~s low-order g-mode periods are typically in the
range $\sigma_{\rm rms}$\,$\lesssim$\,0.3~s \citep[e.g.,][]{bischoff-kim_2014_aa}, 
for a fit precision of $\sigma_{\rm rms}/ P$\,$\lesssim$\,0.3\%.
Lower root-mean-square residuals using ab initio WD models are possible \citep{charpinet_2019_aa,giammichele_2022_aa},
although see \citet{de-geronimo_2019_aa}.

Our finding of relative period shifts of $\Delta P/P$\,$\simeq$\,$\pm$\,2.0\% 
suggests that an astrophysical constraint on the
$^{12}$C$(\alpha,\gamma)^{16}$O reaction rate probability distribution function could, in principle, 
be extractable from the derived period spectrum of observed variable WD.  
Our results can inform future inferences, including those from
machine-learning \citep[e.g.,][]{bellinger_2016_aa},
on the interior mass fraction profiles and the reaction rates that
produce those chemical profiles.

\acknowledgements

We thank James Deboer for sharing the $^{12}$C$(\alpha,\gamma)^{16}$O probability
distribution function, Josiah Schwab for sharing \code{wd\_builder}, 
and Pablo Marchant for sharing \code{mkipp}.
This research is supported by NASA under the Astrophysics Theory Program grant NNH21ZDA001N-ATP, 
by the NSF under grant PHY-1430152 for the 
``Joint Institute for Nuclear Astrophysics - Center for the Evolution of the Elements'', and 
by the NSF under the Software Infrastructure for Sustained Innovation
grants ACI-1663684, ACI-1663688, and ACI-1663696 for the \MESA\ Project.
This research made extensive use of the SAO/NASA Astrophysics Data System (ADS).

\software{
\MESA\ \citep[][\url{https://docs.mesastar.org/}]{paxton_2011_aa,paxton_2013_aa,paxton_2015_aa,paxton_2018_aa,paxton_2019_aa},
\texttt{MESASDK} 20190830 \citep{mesasdk_linux,mesasdk_macos},
\code{wd\_builder} \url{https://github.com/jschwab/wd_builder},
\GYRE\ \citep[][\url{ https://github.com/rhdtownsend/gyre}]{townsend_2013_aa,townsend_2018_aa},
\code{mkipp} \url{https://github.com/orlox/mkipp},
\texttt{matplotlib} \citep{hunter_2007_aa}, and
\texttt{NumPy} \citep{der_walt_2011_aa}.
         }

\clearpage

\bibliographystyle{aasjournal}


\end{document}